\documentclass[aps,prl,twocolumn,showpacs,amsmath,amssymb]{revtex4}
\usepackage{epsfig}
\usepackage{bm}

\newcommand{\be}{\begin{equation}} \newcommand{\ee}{\end{equation}}

\newcommand{\ba}{\begin{eqnarray}}
\newcommand{\ea}{\end{eqnarray}}

\newcommand{\bq}{\begin{equation}}
\newcommand{\eq}{\end{equation}}
\newcommand{\bqa}{\begin{eqnarray}}
\newcommand{\eqa}{\end{eqnarray}}
\newcommand{\ben}{\begin{enumerate}}
\newcommand{\een}{\end{enumerate}}
\newcommand{\bc}{\begin{center}}
\newcommand{\ec}{\end{center}}
\newcommand{\bqb}{\begin{eqnarray*}}
\newcommand{\eqb}{\end{eqnarray*}}

\begin{document}

\title{\vspace{1cm}
Exact Ground State and Finite Size Scaling in a 
\\Supersymmetric Lattice Model
}

\author{M. Beccaria and G. F. De Angelis}

\affiliation{
Dipartimento di Fisica and INFN, Universit\`a di
Lecce \\
Via Arnesano, 73100 Lecce, Italy
}

\begin{abstract}
We study a model of strongly correlated fermions in one dimension with extended N=2 supersymmetry.
The model is related to the spin $S=1/2$ XXZ Heisenberg chain at anisotropy $\Delta=-1/2$ with a real  
magnetic field on the boundary.
We exploit the combinatorial properties of the ground state to determine its exact wave function on 
finite lattices with up to 30 sites.
We compute several correlation functions of the fermionic and spin fields. We discuss the continuum
limit by constructing lattice observables with well defined finite size scaling behavior.
For the fermionic model with periodic boundary conditions we give the emptiness formation probability in closed form.
\end{abstract}


\maketitle

Supersymmetry is well motivated in 
high energy physics where it offers a partial solution to fine-tuning problems and 
improve gauge coupling  unification~\cite{Witten}. 
It also appears in condensed matter models although at a less fundamental level.
Examples are disordered systems~\cite{Efetov} and models of strongly correlated electrons like 
extended Hubbard~\cite{Hubbard} or t-J models~\cite{tJ} where supersymmetry relates fermionic
and bosonic composite operators.

A typical consequence of unbroken supersymmetry is the prediction of the ground state energy.
This is not sufficient to compute the ground state wave function,
the relevant quantity for the calculation of vacuum expectation values. 
It is quite natural to ask whether supersymmetry and the knowledge of the ground state energy 
are useful to this purpose.

In this Letter, we analyze the problem in a recently proposed one dimensional model of itinerant fermions~\cite{Fendley1}
with two supercharges obeying with the Hamiltonian an extended N=2 supersymmetry algebra. 
The knowledge of the ground state on large finite lattices is important to study its continuum limit
where the model is expected to describe a minimal superconformal series.

We shall be interested in boundary effects and thus consider mainly free boundary conditions, see \cite{Fendley1}
for the periodic case.
The model can be mapped to the integrable open XXZ Heisenberg spin $1/2$ chain 
with anisotropy $\Delta=-1/2$ and a suitable real surface magnetic field. 
In principle, Bethe Ansatz techniques could be applied. The supersymmetry inherited from the fermionic model 
should allow to compute Baxter function whose zeros give the Bethe quantum numbers~\cite{Baxter}.
However, the procedure is rather involved with open boundary conditions as discussed in~\cite{Gier}.

Here, we pursue a different approach  starting from the following remarks. 
The XXZ chain at $\Delta=-1/2$ is integrable for a large class of boundary conditions. 
In some specific cases ({\em e.g.} twisted or $\rm U_q(sl(2))$ symmetric ones) several remarkable conjectures
have been claimed about the combinatorial properties of the ground state wave function~\cite{Conjectures,Batchelor}. 
They arise from the relation between the XXZ chain and Temperley-Lieb loop models~\cite{Loop}.

In this Letter, we first show that similar features are present in the fermionic model and in the related XXZ chain with 
surface magnetic field. We then explain how non standard number theoretical methods can be used to obtain 
exact expressions for the ground state wave function of the fermionic model on long chains. 
Finally, we analyze in some details the physical properties and finite size scaling (FSS) behavior of the ground state. 
In particular, we discuss the continuum limit and propose a way to extract scaling fields from the 
fermionic model and the associated XXZ Heisenberg chain.

The model~\cite{Fendley1} is defined on a one dimensional lattice with $L$ sites and free boundary conditions. 
Let $c_i$, $c_i^\dagger$ be spinless fermionic creation annihilation operators with algebra
$\{c_i, c_j^\dagger\} = \delta_{ij}$, $\{c_i, c_j\} = \{c_i^\dagger, c_j^\dagger\} = 0$.
We denote by $N_i$ the set of nearest neighbours of site $i$.
The projector over states with a hard core condition forbidding occupancy around site $i$ is 
${\cal P}_i = \prod_{j\in N_i}(1-n_j)$, with $n_j = c^\dagger_j c_j$.
Let $Q^+$ be the supersymmetry charge $Q^+ = \sum_i c_i^\dagger\ {\cal P}_i$.
and $Q^- = (Q^+)^\dagger$. The operators $Q^\pm$ are nilpotent. The Hamiltonian
\be
\label{Hamiltonian}
H = \{Q^+, Q^-\} = \sum_i\sum_{j\in N_i} {\cal P}_i \ c_i^\dagger c_j \ {\cal P}_j + \sum_i {\cal P}_i ,
\ee
is by construction $Q^\pm$  symmetric. 
We restrict ourselves on the subspace ${\cal H}_{L,F}$ of states with $F$ fermions and no adjacent occupied sites
which is an invariant subspace of the full Fock space under the action of $Q^\pm$.
We work in the basis of simultaneous eigenstates of the number operators $n_i$.
The structure of eigenstates of $H$ follow from supersymmetry. The energy is non negative and 
all energy eigenstates with $E>0$ are  doublets connected by the action of $Q^\pm$. The 
zero energy states are singlets annihilated by $Q^\pm$. They are supersymmetric ground states.
A cohomological analysis~\cite{Fendley1} shows that there is a unique zero energy state for L mod 3 = 0, 2
and none for L mod 3 = 1. In the following we shall consider the case $L=3n$. The unique ground state has then fermion number $F=n$.
The dimension of ${\cal H}_{L,L/3}$ increases rapidly as $L\to\infty$.
It should be clear that the determination of the unique ground state $|\Psi_0\rangle$ is non trivial on large lattices.

At low $L$, the ground state can easily be obtained.
Inspection of the explicit wave functions reveals a remarkable fact. Indeed, the ground state can always be written in the 
very special form 
\be
\label{Ansatz}
|\Psi_0\rangle = |\underbrace{1010\cdots 10}_{2F \ \rm terms}\underbrace{00\cdots 0}_{F\ \rm terms}\rangle + \sum_s x_s |s\rangle ,
\ee
where $x_s\in\mathbb{Z}$ and $s$ runs over all states (diagonal in the occupation number basis) with the exception of 
the single state explicitly written. Lattice parity halves the number of independent coefficients. We used it as a 
consistency check. The integrality property of $\{x_s\}$ is definitely a non trivial Ansatz on the ground state wave function,
and it would be elusive in a Bethe Ansatz approach.
Similar results already appeared in the literature for the XXZ Heisenberg model at $\Delta=-1/2$~\cite{Conjectures,Batchelor}
that is indeed closely related to the present fermionic model.

Let us assume the integrality property (\ref{Ansatz}) as a working hypothesis.
The equation $H|\Psi_0\rangle = 0$ reduces to a linear problem of the special kind
\be
\label{linalg}
A x = b,\qquad A\in \mathbb{Z}^{d\times d}, b\in \mathbb{Z}^{1\times d} ,
\ee
where $x$ is the vector of unknown coefficients $x_s$ and $d = \dim{\cal H}_{L,F}-1$. We know
that this problem admits an integer solution $x\in \mathbb{Z}^{1\times d}$ and it seems reasonable to be able
to find it exactly with modest effort. This is possible due to a well known technique in cryptology.
The problem can be solved by working in the finite field $\mathbb{Z}_p$ of integers modulo a prime $p$. 
To solve Eq.~(\ref{linalg}), we choose a large prime $p$ and first determine a solution modulo $p$
by the Lanczos algorithm over finite fields~\cite{LaMacchia}.
We build the sequences $\{b_i\}_{0\le i\le d}$ and $\{c_i\}_{1\le i\le d}$ where the initial values are
\be
\label{Lanczos1}
b_0=b,\quad c_1 = Ab,\quad b_1 = c_1-\frac{c_1^2}{b\cdot c_1}b ,
\ee
and the sequences are generated by iterating
\ba
c_{i+1} &=& A b_i, \\
b_{i+1} &=& c_{i+1}-\frac{c^2_{i+1}}{b_i\cdot c_{i+1}} b_i -\frac{c_i\cdot c_{i+1}}{c_i\cdot b_{i-1}}b_{i-1} .
\ea
The solution is finally obtained as
\be
\label{Lanczos2}
x = \sum_{i=0}^{d-1}\frac{b_i\cdot b}{b_i\cdot c_{i+1}}b_i .
\ee
In Eqs.~(\ref{Lanczos1}-\ref{Lanczos2}), all arithmetic operations, in particular  divisions,  are done in $\mathbb{Z}_p$.
The above algorithm exploits the sparsity of $A$, and is not memory expensive since requires a storage ${\cal O}(d)$. 
After a certain number of solutions modulo (large) primes $\{p_i\}$ have been found,
they are combined together to give the exact solution. This is done by applying the Chinese Remainder Theorem~\cite{LeVeque}.
By the above procedure, we have determined the exact ground state for $L=3n \le 30$ ($\dim{\cal H}_{30, 10} = 352716$). 
The complete expressions ({\em i.e.} the sequences $\{x_s\}$) are available on request. 

\begin{figure}[htb]
\vskip 0.7cm
\begin{center}
\leavevmode
\psfig{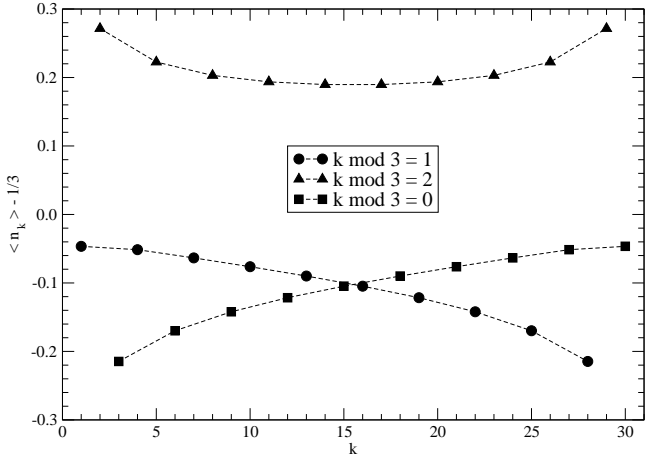}
\vspace{0.1cm}
\caption{$\mathbb{Z}_3$ structure of the expectation $\langle n_k\rangle$. Dashed lines connect branches with the same
value of k mod 3.}
\label{fig:occupation1}
\end{center}
\end{figure}

We now illustrate the main physical properties of the results. For any observable $\cal O$
we denote $\langle {\cal O}\rangle\equiv \langle \Psi_0|{\cal O}| \Psi_0\rangle$. 
Several remarkable combinatorial features appear immediately
leading to conjectures in the spirit of~\cite{Conjectures}. We find that the dominant state in the ground state is 
always $|\tau\rangle = |010 010 010 \cdots\rangle$ with coefficient 
$\max |x_s| = x_{|\tau\rangle} = N_8(2L/3+2)$ where $N_8(2n) = \prod_{k=1}^{n-1}(3k+1)(2k)!(6k)!/((4k)!(4k+1)!)$ is the number of cyclically 
symmetric transpose complement plane partitions~\cite{Batchelor}. Also, the squared norm in our normalization is 
$1+\sum_s x_s^2 = N_8(2L/3+2) A_V(2L/3+3)$,
where $A_V(2n+1) = 2^{-n}\prod_{k=1}^n (6k-2)!(2k-1)!/((4k-1)!(4k-2)!)$ is the number of vertically symmetric alternating sign matrices
We checked these expressions for all $L\le 30$. 
Similar conjectures can be claimed on correlation functions, as in the recent work~\cite{Correlation}.
A simple example is the expectation of the potential energy $\langle\sum_i {\cal P}_i\rangle$.
The operator $(1/L) \sum_i {\cal P}_i$ is diagonal on states $|{\bf n}\rangle$ and is minimum on $|\tau\rangle$ where it attains the value $F/L=1/3$.
Our data are consistent with the simple formula
\be
\label{FermionPotential}
\frac{1}{L}\langle\sum_i {\cal P}_i\rangle = \frac{1}{3}\ \frac{5L+21}{4L+15} .
\ee
The asymptotic value is $5/12$, slightly larger than 1/3, due to the subdominant states  $|{\bf n}\rangle\neq |\tau\rangle$
appearing in $|\Psi_0\rangle$. Another example is the fermion density on the boundary   $\langle n_1\rangle$. We find
\be
\label{FermionBoundary}
\langle n_1 \rangle = \frac{L(10L+33)}{2(4L+9)(4L+15)} .
\ee
The asymptotic value $5/16$ is smaller than $1/3$, a plausible fact since the dominant state $|\tau\rangle$ has no
fermions on the lattice boundary. 
We now examine the FSS behavior of simple correlation functions. In Fig.~(\ref{fig:occupation1})
\noindent
we show the expectation value $\langle n_k\rangle-1/3$ on the L=30 lattice. There is a clear $\mathbb{Z}_3$
substructure similar to that observed in the phase diagram of  bosonic models with hard-core
repulsion~\cite{Fendley4}. The average number of fermions in the three branches can be defined
as $F_k = \sum_{i, i\ {\rm mod}\ 3 = k}\langle n_i\rangle$. We have $F_0 = F_1$ and $F_0+F_1+F_2 = F$. Asymptotically, 
at large $F = L/3$, the upper branch has the asymptotic value
$F_2 = \frac{1}{2}(F+1) + {\cal O}(F^{-1})$.
The three sublattices containing sites $k$ with fixed $k$ mod 3 appear to reconstruct 
a smooth curve for $\langle n_k\rangle$ as $L$ increases.
The $\mathbb{Z}_3$ structure can be exploited
to construct scaling fields in the continuum limit.
\begin{figure}[htb]
\vskip 0.7cm
\begin{center}
\leavevmode
\psfig{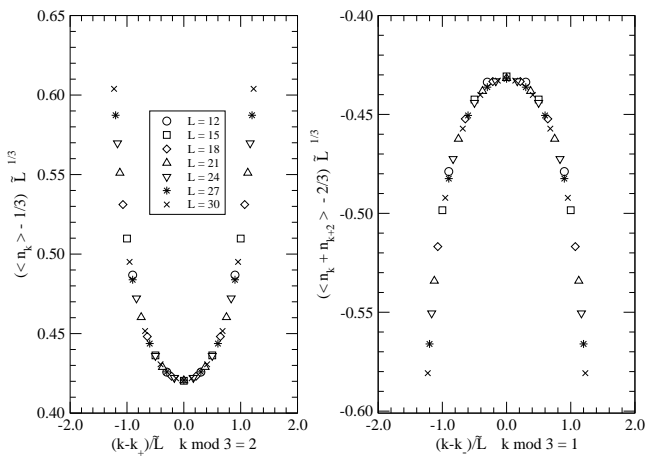}
\vspace{0.1cm}
\caption{FSS of the occupation number $\langle n_k\rangle$.}
\label{fig:occupation2}
\end{center}
\end{figure}
\noindent
Let us define the effective length $\widetilde L = L/3+1$, and set $k_\pm = (L\pm 1)/2$. 
Our data for $\langle n_k\rangle$ suggest to test the following FSS laws
\be
\label{FSS1}
\langle n_k \rangle - 1/3 = f_+\left((k- k_+)/{\widetilde L}\right) {\widetilde L}^{-\nu},\ \ k\ \mbox{mod}\ 3 = 2 ,
\ee
$$
\label{FSS2}
\langle n_k + n_{k+2} \rangle - 2/3 = f_-\left((k- k_-)/{\widetilde L}\right) {\widetilde L}^{-\nu'},\ \ k\ \mbox{mod}\ 3 = 1 ,
$$
where $\nu$ and $\nu'$ are unknown exponents. The best collapse of data at different $L$ is obtained with 
$\nu = \nu' = 0.33(2)$ and is shown in Fig.~(\ref{fig:occupation2}).
Notice that for $k\ \mbox{mod}\ 3 = 1$,
the separate $\langle n_k \rangle$ and $\langle  n_{k+2} \rangle$ form a parity doublet. They should be
related in the continuum limit to the left and right moving parts of a single field.
The value of $\nu$, $\nu'$ is close to $1/3$. In principle, this information is 
useful in the effort of identifying the proposed scaling fields with the operator content of 
candidate  superconformal field theories describing the continuum limit of the model.
The next simple observable built with local fermionic fields is the density-density (connected) 
correlation function 
$
G_{i,j} = \langle n_i n_j \rangle - \langle n_i\rangle\langle n_j\rangle .
$
\begin{figure}[htb]
\vskip 0.7cm
\begin{center}
\leavevmode
\psfig{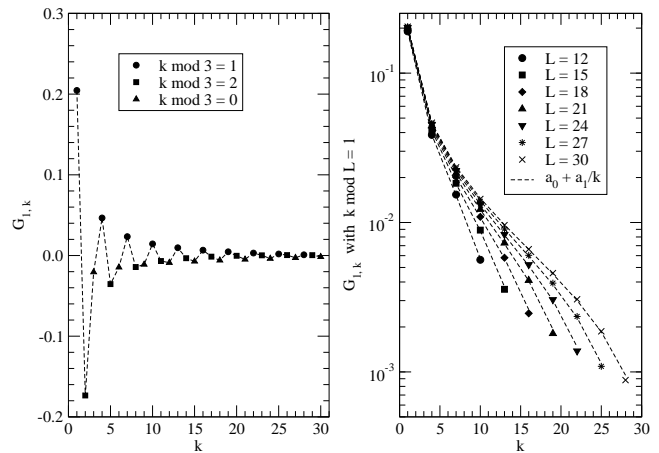}
\vspace{0.2cm}
\caption{Boundary and bulk pair correlation function at L=30. Algebraic decay of the $k$ mod 3 = 1 component
of the boundary correlation.}
\label{fig:pair}
\end{center}
\end{figure}
\noindent
On the left side of Fig.~(\ref{fig:pair}) we show the boundary correlation
function $G_{1,k}$ on the L=30 lattice. Again the $\mathbb{Z}_3$ structure is evident. On the right side, 
we fit $G_{1,k}$ on lattices of various sizes with the simple form $a_0+a_1/k$ with good agreement. 
The term $a_0$ is a small size effect decreasing with $L$. A detailed scaling analysis shall be reported elsewhere.

\begin{figure}[htb]
\vskip 0.7cm
\begin{center}
\leavevmode
\psfig{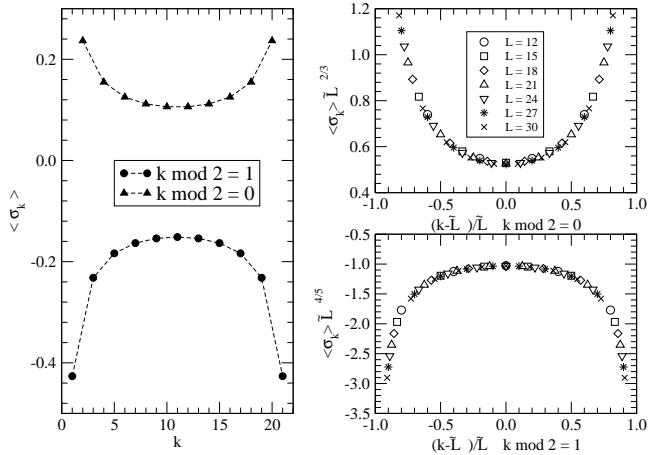}
\vspace{0.2cm}
\caption{(left) $\mathbb{Z}_2$ structure, (right) FSS of $\langle\sigma_k\rangle$.}
\label{fig:spin}
\end{center}
\end{figure}

The fermionic model can be mapped quasi-locally to the XXZ spin chain at anisotropy $\Delta=-1/2$
with a specific real surface magnetic field~\cite{Fendley3}. The structure of
Bethe equations in the two models is closely related, but the fermionic and spin correlation functions 
can be quite different. Following~\cite{Fendley3}, we map a fermionic configuration $\{{\bf n}\}$ 
to a well defined spin one $\{{\bf \sigma}\}$ with the dichotomic variable $\sigma_k = \pm 1$. The map is such that the spin chain 
has length $2L/3+1$. The expectation value of the local spin is shown on the left side of Fig.~(\ref{fig:spin}).
\noindent
The structure is now $\mathbb{Z}_3\to \mathbb{Z}_2$ and $\langle \sigma_k\rangle$ can be split into two 
independent channels with $k$ mod 2 = 0,1.
Eqs.~(\ref{FermionPotential},\ref{FermionBoundary}) give the boundary 
spin $\langle \sigma_1\rangle=2\langle n_1\rangle-1$ and the numbers of pairs of spins up or down 
\be
\langle N_{\uparrow\uparrow}\rangle = \frac{L(L-3)}{3(4L+9)},\ 
\langle N_{\downarrow\downarrow}\rangle = \frac{L(L+6)}{3(4L+15)}. 
\ee
Again, it is possible to separate the two components in order to build well defined scaling 
fields. We test
\be
\langle \sigma_k \rangle = g_+\left((k- \widetilde L)/{\widetilde L}\right) {\widetilde L}^{-\beta},\ \ k\ \mbox{mod}\ 2 = 0 ,
\ee
\be
\langle \sigma_k \rangle = g_-\left((k- \widetilde L)/{\widetilde L}\right) {\widetilde L}^{-\beta'},\ \ k\ \mbox{mod}\ 2 = 1 .
\ee
The best values of the exponents are $\beta=0.66(1)$, $\beta'=0.80(1)$. The accuracy of the corresponding FSS laws 
is shown on the right side of Fig.~(\ref{fig:spin}). Notice that $\beta$, $\beta'$ are quite close to the
simple rationals $2/3$ and $4/5$. This suggests again possible simple identification of these scaling fields
with superconformal operators. 

We want to stress at this point that the same methods can be applied to any model with integrality properties like (\ref{Ansatz}). 
In particular,  a suitable version of (\ref{Ansatz}) holds in the model  (\ref{Hamiltonian}) with periodic boundary conditions.
To give an example of the efficacy of our number theoretical techniques in this context, we have 
computed the unique supersymmetric ground state for the model with periodic boundary conditions and 
$L=3F+1$, up to $L=28$. A relevant non-trivial observable is the emptyness formation probability
$
E_k = \langle\prod_{i=1}^k (1-n_i)\rangle.
$
The first values are fixed from translation invariance and the constraint $n_i n_{i+1}=0$
which defines ${\cal H}_{L,F}$
\be
E_1 = \frac{2F+1}{3F+1},\qquad E_2 = \frac{F+1}{3F+1} .
\ee
For $k>2$ we checked the validity of the relation
\be
\label{wow}
E_k = E_{k-1}\ \frac{(k-2)!(3k-5)!}{(2k-3)!(2k-4)!}\ \prod_{\nu=3-k}^{k-1}\frac{F+\nu}{2F+\nu} .
\ee
We guessed the very specific form of Eq.~(\ref{wow}) from similar conjectured relations proposed for 
the finite XXZ model with twisted boundary conditions~\cite{Conjectures}.
The thermodynamical limit $F\to\infty$ is 
\be
\varepsilon_k \equiv \lim_{F\to\infty} E_k = \frac{1}{24}A(k-1) 2^{-(k-3)(k+1)},
\ee
where $A(n)$ is the number of alternating sign matrices
$A(n) = \prod_{k=0}^{n-1} (3k+1)!/(n+k)!$~\cite{Batchelor}. 
From Stirling's expansion, the asymptotic behaviour of this expression is 
\be
\varepsilon_{k+1} \sim 2/3\ c \left(\sqrt{3}/2\right)^{3k^2} \ k^{-5/36},
\ee
where the constant $c$ takes the form \cite{Conjectures}
\be
c = \exp \int_0^\infty \left(\frac{5e^{-t}}{36}-\frac{\sinh(5t/12)\sinh(t/12)}{\sinh^2(t/2)}\right)\frac{dt}{t}
\ee
To conclude, we have shown that the knowledge of the ground state $|\Psi_0\rangle$  on moderately large finite lattices allows 
a precise FSS analysis of correlation functions guiding the detailed 
identification of the continuum limit
of the model (\ref{Hamiltonian}). The exact $|\Psi_0\rangle$ is also an  effective heuristic tool to derive 
closed expressions for correlation functions.
We acknowledge conversations with P. Fendley and  J. De Boer.

\end{document}